\documentclass[a4paper]{scrartcl}

\pdfoutput=1

\usepackage[T1]{fontenc}
\usepackage[utf8]{inputenc}
\usepackage[english]{babel}
\usepackage{lmodern}
\usepackage{graphics}
\usepackage{float}
\usepackage{bbold}
\usepackage[colorlinks=true,linkcolor=blue]{hyperref}
\usepackage{float}
\usepackage{framed}
\usepackage[table]{xcolor}

\usepackage{listings}
\definecolor{mygreen}{RGB}{28,172,0} 
\definecolor{mylilas}{RGB}{170,55,241}

\usepackage{url}

\usepackage{draftwatermark} 
\SetWatermarkText{Preprint} 
\SetWatermarkScale{1}       
\SetWatermarkLightness{0.9} 

\usepackage{pgfplots}

 \usepgfplotslibrary{external}
 \tikzexternaldisable

\renewcommand{\url}[1]{\href{http://#1}{\texttt{#1}}}

\newcounter{saak}

\newcounter{fehr}

\newcounter{himpe}

\newcounter{heiland}

\begin{document}

\title{Best Practices for Replicability, Reproducibility and Reusability of Computer-Based Experiments Exemplified by Model Reduction Software}

\author{
J\"org Fehr\thanks{Institute of Engineering and Computational Mechanics at the University of Stuttgart, Pfaffenwaldring 9, D-70569 Stuttgart, Germany
(\href{mailto:joerg.fehr@itm.uni-stuttgart.de}{\nolinkurl{joerg.fehr@itm.uni-stuttgart.de}})} ~
Jan Heiland\thanks{Computational Methods in Systems and Control Theory Group at the Max Planck Institute for Dynamics of Complex Technical Systems, Sandtorstra{\ss}e~1, D-39106 Magdeburg, Germany
(\href{mailto:heiland@mpi-magdeburg.mpg.de}{\nolinkurl{heiland@mpi-magdeburg.mpg.de}})} ~
Christian Himpe\thanks{Institute for Computational and Applied Mathematics at the University of M\"unster, Einsteinstrasse~62, D-48149 M\"unster, Germany
(\href{mailto:christian.himpe@uni-muenster.de}{\nolinkurl{christian.himpe@uni-muenster.de}})} ~
Jens Saak\thanks{Computational Methods in Systems and Control Theory Group at the Max Planck Institute for Dynamics of Complex Technical Systems, Sandtorstra{\ss}e~1, D-39106 Magdeburg, Germany
(\href{mailto:saak@mpi-magdeburg.mpg.de}{\nolinkurl{saak@mpi-magdeburg.mpg.de}})}
}

\date{}

\renewcommand{\ttdefault}{pcr}
\lstset{language=Matlab,%
	basicstyle=\ttfamily\footnotesize,%
    breaklines=false,%
    morekeywords={matlab2tikz},
    keywordstyle=\color{blue},%
    morekeywords=[2]{1}, keywordstyle=[2]{\color{black}},
    identifierstyle=\color{black},%
    stringstyle=\color{mylilas},
    commentstyle=\color{mygreen},%
    showstringspaces=false,
    numbers=left,%
    numberstyle={\tiny \color{black}},
    numbersep=5pt, 
}

\maketitle

\begin{abstract}
Over the recent years the importance of numerical experiments has gradually been more recognized.
Nonetheless, sufficient documentation of how computational results have been obtained is often not available.
Especially in the scientific computing and applied mathematics domain this is crucial, since numerical experiments are usually employed to verify the proposed hypothesis in a publication.
This work aims to propose standards and best practices for the setup and publication of numerical experiments. Naturally, this amounts to a guideline for development, maintenance, and publication of numerical research software.
Such a primer will enable the replicability and reproducibility of computer-based experiments and published results and also promote the reusability of the associated software.
\end{abstract}



\setcounter{tocdepth}{1}


\section{Introduction}

In a publication in the fields of applied mathematics, numerical analysis, and scientific computing, a Computer-Based Experiment (CBEx) or its results can be of different value.
If a work contains strong and generally valid analytical findings, a CBEx may not be needed or is just used to affirm a valid fact by some concrete numerical results.
On the other hand, if the considered problem is very complex or very specific, a practical example might be necessary to justify a possibly wild combination of analytical estimates, intuitive assumptions, or heuristics.
 In the extreme case, there might be no analytical reasoning at all and the whole research contribution bases on CBEx.

 One may well say, that with increasing complexity of the considered problems and with increasing computation capabilities, both the need for and the opportunity to provide a valid CBEx to a scientific work has grown.

 Exemplarily, this general observation can be illustrated by comparing three papers from 1971, 1986, and 2010, which introduced nowadays commonly applied numerical methods. 
In Nitsche's 1971 paper \cite{Nit71} on a new variational approach to elliptic PDEs with non-homogeneous Dirichlet conditions,
there is no numerical experiment reported.
Then, in the important paper \cite{SaaS86} by Saad and Schultz on the \emph{GMRES} algorithm from 1986, two out of 14 pages are devoted to numerical experiments.
Finally, the paper \cite{ChaS10} on \emph{DEIM} by Chataranbutat and Sorensen in 2010, consists of more than 30\% of numerical examples or reasonings based on numerical experiments.

 Summing up, we assess that the value of a CBEx has risen significantly in comparison to analytical results over the last decades.
However, the high standards on analytical findings, namely the requirement of a concise and comprehensible and traceable derivation and documentation, seems not equally adapted to numerical experiments and results in the scientific literature, cf. LeVeque's article on \emph{Top Ten Reasons to Not Share Your Code (and why you should anyway)} \cite{LeV13}.

With the ever growing sophistication of the numerical simulations, a CBEx in the field of mathematics has more and more changed its nature.
From a rather deterministic mathematical exercise on a computer (which is still remembered in the terms \emph{numerics} and to some extend in \emph{numerical} referring more to numerology than to floating point operations) towards a scientific experiment with inevitable uncertainties coming, e.g. from rounding errors or changing software and hardware environments.
Thus, a CBEx should be seen in analogy with experiments from natural sciences with the numerical result corresponding to the observation of the experiment and with the hard- and software corresponding to the methods that were used to obtain the observations like the experimental setup, the design of the tests, the used statistics, or the choice of the samples. 
 
Once an experiment has been established, the question of \emph{reproducibility} arises, since only an experiment and its obtained observations which can be reproduced, is seen to give valid and reliable insights that can serve as the base for further research.
This principle seems broadly accepted since long, and it has found its formulation in Popper's work \emph{Logik der Wissenschaften} from 1935, later translated into English, with the formulation ``I only demand that every such statement must be capable of being tested;
or in other words, I refuse to accept the view that there are statements in science which we have, resignedly, to accept as true merely because it does not seem possible, for logical reasons, to test them.'', cf. \cite[Ch.~1.8]{Pop02}.
Note that the demand of testability of the hypothesis does not include a \emph{truth} value as it is implicated by the reproducibility of an experiment.
However, as Popper states, an unreproducible singular discovery would not be published by a researcher, since ``the ‘discovery’ would be only too soon rejected as chimerical, simply because attempts to test it would lead to negative results.'', cf. \cite[Ch.~1.8]{Pop02}

Reproducibility is commonly accepted as a necessary condition for good scientific practice, and it's absence in some prominent works but also in a statistically significant number of journal publications that has been detected in recent years in, e.g., medicine \cite{ErrIGetal14}, psychology \cite{JohLP12}, and computer science \cite{ColPW14} has shaped the term of the \emph{reproducibility crisis} that has been broadly covered in scientific, public, and social media\footnote{\href{http://newyorker.com/tech/elements/the-crisis-in-social-psychology-that-isnt}{\texttt{newyorker.com/tech/elements/the-crisis-in-social-psychology-that-isnt}}}
\footnote{\href{http://bjoern.brembs.net/2016/02/earning-credibility-in-post-factual-science}{\texttt{bjoern.brembs.net/2016/02/earning-credibility-in-post-factual-science}}}.

The general concept of reproducibility has been taken up in computer-based research in the 90s \cite{BucD95} and adapted to the comparatively deterministic nature of software and its ability to easily enable the ``open exchange of data, procedures and materials'', as it was phrased in a code of ethics and values of the American Physical Society\footnote{\href{http:\\www.aps.org/policy/statements/99_6.cfm}{\texttt{aps.org/policy/statements/99\_6.cfm}}}.
In this time, the term \emph{reproducible research} \cite{fomel09} was shaped and often referred to computational environments that allowed for simply transferring to and rerunning the experiments on different computers; see \cite{Mar16} for an example in the field of archeology and for references. 
 
It is also in the nature of software that it can be duplicated and dissected so that not only the results but also parts of the methods itself can serve as the base of new experiments, which is meant by \emph{reusability}. 

In this work, we adapt notions related to Replicability, Reproducibility and Reusability (\emph{RRR}) as they are relevant for CBEx from first principles.
We describe conditions for their implementation in research and publications that are general enough to meet particular needs of projects as well as habits of the researchers.
To find the balance between a reliable framework and openness towards common practices, we add sections with concrete suggestions -- a \emph{best practice guide}.

In this contribution the details on code and data layout or licensing and associated copyright issues are not covered; work on these topics can be found for example in \cite{stodden09a} and \cite{stodden14} respectively.
Also, for completeness we mention that our work is about the way how CBEx are conducted and documented.
Hence, the principles considered here are to be distinguished from approaches that try to validate numerical results like the notion of \emph{Verification and Validation}\footnote{\href{http://sciencenode.org/feature/why-should-i-believe-your-hpc-research-.php}{\texttt{sciencenode.org/feature/why-should-i-believe-your-hpc-research-.php}}}.

 Overall, this work aims to: \emph{Make CBEx replicable in its basic definition and use the potential of software to enable easy reproducibility and even reusability.}

\subsection{Prior Work and State of the Discourse}
 The discrepancy between the potential of CBEx to be easily made RRR and the widespread lack of RRR in CBEx in the scientific literature has stimulated various initiatives and theoretical work on the implementation of RRR principles in scientific computing. We list but a few of the most recent publications:

The discussion on opening scientific source codes has been more noticeable in the recent years.
For example in \textit{Nature}, arguments against open source are refuted \cite{barnes10},
more accurate results are predicted \cite{merali10}, partial opened codes are discussed \cite{ince12}, and a code availability section is suggested \cite{codeshare,ctrlalt}.
In \textit{Science} not only the opening and review of research codes is discussed \cite{joppa13,sliz13,joppa13a} but it is required by the editorial policies that:
``\textit{All computer codes involved in the creation or analysis of data must also be available to any reader of Science}''.
Also mathematical organizations are discussing open scientific codes, examples are
\textit{AMS} on the maintainability and necessity of open code accompanying publications \cite{joyner07},
\textit{ACM} on advantages and disadvantages of releasing the scientific codes \cite{mccafferty10},
and \textit{SIAM} on a publication of codes by default and attributable credit \cite{bangerth14}.

Several publications describe abstract software engineering and collaborative development techniques.
In \cite{kelly09} basic practices for scientific software development are distilled,
while in \cite{heroux09} software management principles are explained.
A set of rules, devised in \cite{prlic12}, is concerned with the code development but also the user-developer interaction. 
And the best practices in \cite{wilson14} summarize code development fundamentals.
General recommendations for reproducibility for CBEx are also given in \cite{bailey16}.
Furthermore, the practical reproduction of research results themselves is discussed as in \cite{mesnard16}.

Lastly, we note that various initiatives have been started to promote certain standards in CBEx.
Foremost, the \emph{Science Code Manifesto}\footnote{\href{http://sciencecodemanifesto.org}{\texttt{sciencecodemanifesto.org}}} states five principles (Code, Copyright, Citation, Credit, Curation) for the handling of research software to improve its use in science. 
The \emph{Recomputation Manifesto}\footnote{\href{http://recomputation.org}{\texttt{recomputation.org}}} \cite{gent13} also formulates rules to facilitate the repeatable realization of CBEx.

\subsection{Outline}
This introductory discussion is followed by a more refined analysis of replicability, reproducibility and reusability in \href{sc:rrr}{Section~\ref*{sc:rrr}}.
In \href{sc:cas}{Section~\ref*{sc:cas}} a technique to document code availability is described.
\href{sc:cog}{Section~\ref*{sc:cog}} summarizes high-level considerations to facilitate RRR, while a minimal documentation for scientific codes and research software is proposed in \href{sc:definfo}{Section~\ref*{sc:definfo}}.
Finally, a sample software project is presented to illustrate the practical implementation of the herein suggested best practices.

\section{The Three ``R''s of Open Science}\label{sc:rrr}
In this section, taking up the ideas of \cite{vitek11}, we give a definition of the frequently used terms \textbf{Replicability}, \textbf{Reproducibility}, and \textbf{Reusability} and discuss how these basic scientific principles apply for assessing scientific software.

The distinct notions of \textbf{Replicability} and \textbf{Reproducibility} are used to qualify research in all fields of science in which experiments play a role, cf., e.g. \cite{vaux12} with a background in biology, \cite{opensciencecollab15} from psychology, or \cite{easterbrook14,fomel09} focusing on scientific computing.

In short, replicability refers to a repetition of the experiment with the same results by the same observers in the same environment; reproducibility refers to an independent repetition of the experiment and its outcomes in different circumstances. 

Reproducibility points to a certain reliability of both the findings of the experiment and the procedure that was used to obtain the results \cite{krishnamurthi15}. 
Once reliability of a method is established, one can address reusability as the property that enables the use of the method for different setups and different purposes.

Note that these characteristics should be considered nested, which means reproducibility implies replicability and reusability require reproducibility.

In what follows, we extend, specify, and adapt these general notions to the case of scientific software and numerical simulations.

\subsection{Replicability}\label{sc:replicability}
The attribute \textbf{Replicability} describes the ability to repeat a CBEx and to come to the same (in a numerical sense) results.
Sometimes the equivalent term \textbf{Repeatability} is used for this experimental property. 
Replicability requires some basic documentation on how to run the software (described in \href{sc:reusereq}{Section~\ref*{sc:reusereq}}) to obtain replicable results.

Replicability, in turn, is a basic requirement of reliable software as well as of its result as it shows a certain robustness of the procedure against statistical influences and bias of the observer.
Also, a replication can serve as a benchmark to which new methods can be compared as pointed out in \cite{vitek11}.

\subsection{Reproducibility}\label{sc:reproducibility} 
In its native definition, \textbf{Reproducibility} of a CBEx means that it can be repeated by a different researcher in a different computer environment.
This can be assured, first, through a documentation that provides enough mathematical and technical detail to set up the CBEx that will provide comparable results, including the software implementation of algorithms;
second, through the distribution of a software capable of producing the results on a large variety of machines,
or third, any combination of these two extrema -- sufficient documentation and available software.
If the CBEx depends on hardware, e.g. if runtime is measured, then for reproducibility the hardware needs to be available or sufficiently well documented.

\subsection{Reusability}\label{sc:reusability}
In the sphere of CBEx, \textbf{Reusability} refers to the possibility to reuse the software or parts thereof for different purposes, in different environments, and by researchers other than the original authors.
In particular, \textbf{Reusability} enables the utilization of the test setup or parts of it for other experiments or related applications.
Although theoretically, any bit of a software can be reused for different purposes, here, \textbf{Reusability} applies only for reproducible parts, since a building block of a CBEx that does not define reproducible or even replicable outcomes cannot be reused for a replicable or reproducible CBEx.


\section{Code Availability Section}\label{sc:cas}
Even though availability of the source code associated to a CBEx is not a requirement for replicability and reproducibility (see \href{sc:cog}{Section~\ref*{sc:cog}}),
it is essential to open the CBEx to peer scrutiny and highly recommended by the authors.
The availability of the source code itself is necessary for reusability and unconditionally desirable for reproducibility.
This section makes the case for a \textbf{Code Availability Section} as introduced by \textsc{Nature} \cite{natureXX,codeshare,ctrlalt}.
Such a section should by default be included in any publication presenting numerical results like a ``Materials and Methods'' section in other sciences, and should state if the utilized code is available and if not for what reason, i.e. third-party licenses, non-disclosure agreements, trade secrets, or the thought of keeping competitive advantages.

Different code availability models exist, which will be listed and shortly commented in the following.
\paragraph{Open source code, published under a public license} Compare, e.g., the iterative rational Krylov algorithm (IRKA) example in \href{sc:min_max_example}{Section~\ref*{sc:min_max_example}}.
This procedure is probably preferred by most scientists and for some people the only way to do proper science, compare, e.g., \cite{ince12}. Referees and interested readers can check if the code fulfills the necessary requirements for reproducibility and they can modify and use the code for their own purpose.
There are multiple possibilities how access to the code can be gained.
Nowadays, a common and widely used procedure is the provisioning of source code via a publicly readable revision control repository located on a private server\footnote{e.g., \url{gitlab.com}} or a third-party service provider\footnote{e.g., \url{github.com}, \url{bitbucket.org}}.
Alternatively, a download from a collection such as \textbf{netlib}\footnote{\url{netlib.org}} can be provided.
A shining example for best practice in the field of open source code in combination with reproducible experiments is the \textit{Image Processing On Line} (IPol)~Journal~\cite{IPOL}.
In this journal each article is supplemented with its source code, with an online demonstration facility and an archive of experiments. Furthermore, the text, as well as source code, are peer-reviewed.

\paragraph{Closed source, software available under a non-public license}
This less desirable option gives readers and reviewers the opportunity to check, e.g. if the proposed numerical procedure / experiments work with their own data, given a license is available.
Often, the source code is encoded or obfuscated to protect intellectual properties, which then allows a replication but not a comprehension of results.
\textsc{Matlab} code, as an example of an interpreted language, can be encoded via the \texttt{pcode} command or compiled into a binary format.
However, as stated since Matlab Version 2014b \cite{Matlab14} ``The \texttt{pcode} function obfuscates the code but does not encrypt it.
While the content in a \texttt{.p} file is difficult to understand, it should not be considered secure.''
For programs written in a compiled language, such as \texttt{C++}, only executables or runtime libraries are provided.
Hence, for trust reasons it is important, that the software has a-priori passed through a strictly documented verification \& validation procedure. 
By providing and hosting the source via a version control repository (see \href{sc:reuse2}{Section~\ref*{sc:reuse2}}) it is possible to provide certain people, i.e. the reviewers, with access to the source code upon request.
Alternatively, the source code may be provided directly to an eligible user via physical data volumes, or direct file transfers.

\paragraph{Software as a Service (SaaS)} The availability of web access to computer programs or computer resources is an emerging strategy.
This approach can also be used to enable interested users or reviewers to use the developed software as a service, e.g.\ to test if the program runs with their own, respectively modified input data. Therefore, SaaS offers many advantages such as read without copying the source code restriction, time-limited access for users, third-party software dependencies can be resolved, new licensing schemes and so on.
It should be noted that, while SaaS enables the use of a CBEx, it does not allow a dissection at a source code level.

\paragraph{Non-available code} The last and the most undesirable option is the non-availability option.
The source code, computer program or required third party software is not available or purchasable to the interested reader.
A review is hardly, possible and the proposed numerical scheme or ideas need to be written in great detail, so reproducibility of the work is possible in a different environment. 

\medskip
A sample \textbf{Code Availability Section} is enclosed in \href{fig:cas}{Figure~\ref*{fig:cas}}.
The linked source code archive should ideally be uniquely identified by a Digital Object Identifier\footnote{\url{doi.org}} (DOI) which can be obtained for software releases for example from \textbf{Zenodo}\footnote{\url{zenodo.org}} for scientific codes.
Alternatively, the source code can be enclosed in the supplemental materials or deposited at some stable location.

\begin{figure}[h!]
  \begin{framed}
    \textbf{Code Availability / Licensing Option} \\
    The source code of the implementations used to compute the presented results can be obtained from: \\
    \begin{center}
    doi:XXXXXXX/XXXXXXXX and is authored by: XXXX, XXXX 
    \end{center}
    ~\\ Please contact XXXXX for licensing information
  \end{framed}
  \caption{Sample Code Availability Section.}
  \label{fig:cas}
\end{figure}

Even though a simple statement on the (non-)availability of the source code does neither improve the review process nor the reproducibility (in the sense of \href{sc:reproducibility}{Section~\ref*{sc:reproducibility}}), it can at least facilitate replicability through its assurance by the authors.
Furthermore, it could be noted if the referees had access to the implementation during the peer review process.

Moreover, due to the important role of computational results, not only in numerical analysis but also in many other sciences, this measure contributes to the basic idea of verifiability in science.
If the source code is made available, as a part of the publication, on the one hand, effort invested into an openly available software implementation is made visible and, on the other hand, compels authors to comment on means of the experimental setup.
Lastly, a mandatory code availability section raises awareness for RRR.

\section{Code Guidelines}\label{sc:cog}
In this section, based on the previous definitions of replicability, reproducibility, and reusability,
guidelines for the design, documentation, or publication of CBEx and research software are summarized.
The foundation for these guidelines is the interrelation of RRR: reusability implies reproducibility which implies replicability;
and are composed of mandatory requirements and optional recommendations. 
 \textbf{Requirements} are limited to the minimal extent necessary while \textbf{recommendations} enable a practical and comfortable realization of the replication, reproduction, or reuse.
The interdependence of the requirements and recommendations is to be understood as follows:
A requirement for replicability is also a requirement for reproducibility and similarly, a requirement for reproducibility is also a requirement for reusability.
The recommendations are optional but strongly encouraged, yet have no dependence on previous recommendations.

\begin{figure}
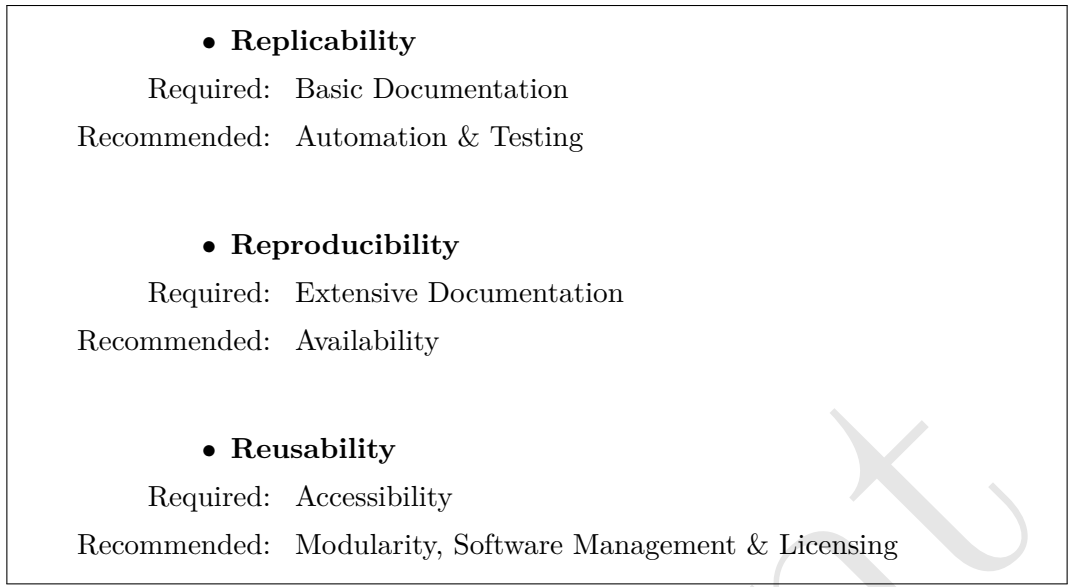
\centering

\begin{minipage}{.95\textwidth}
\begin{framed}\centering
\begin{minipage}{.75\textwidth}
 \begin{itemize}
  \item[$\bullet$] \textbf{Replicability}
   \begin{itemize}
    \item[Required:~] Basic Documentation
    \item[Recommended:~] Automation \& Testing
   \end{itemize}
~
  \item[$\bullet$] \textbf{Reproducibility}
   \begin{itemize}
    \item[Required:~] Extensive Documentation
    \item[Recommended:~] Availability
   \end{itemize}
~
  \item[$\bullet$] \textbf{Reusability}
   \begin{itemize}
    \item[Required:~] Accessibility
    \item[Recommended:~] Modularity, Software Management \& Licensing
   \end{itemize}

 \end{itemize}
\end{minipage}
\end{framed}
\end{minipage}

 \caption{Coding guidelines overview.}
 \label{tb:guidelines}

\end{figure}

We will use the term ``source code archive'' to refer to the set of source code,
build instructions (such as a \texttt{makefile}), configuration files and input data\footnote{The source code archive may also include resulting data sets from the authors experiments.}.
For a summary of the following guidelines see \href{tb:guidelines}{Figure~\ref*{tb:guidelines}}.

\subsection{Replicability Requirement: Basic Documentation}\label{sec:replic-req}
A fundamental requirement for replicability is a \textbf{basic documentation},
which encompasses instructions on how to generate an executable binary program in case of a compiled language,
and a description on how to run the program to obtain the results to be replicated (see also \href{sc:definfo}{Section~\ref*{sc:definfo}}).
This documentation is crucial to an experiment's replication as it defines the technical implementation and ensures the practical repetition of the experiment.

Often, the numerically computed results are further processed to facilitate interpretation, for example by a visualization.
A documentation of the evaluation of these results, descriptively or algorithmically,
is needed to allow for replication, not only of the computational results, but also of their evaluation.

\subsection{Replicability Recommendation: Automation and Testing}
The automation of the experiment enables the easy and reliable check for replicability of a CBEx.
This typically means that a single or multiple scripts automatically
prepare and run the experiment as well as the post-processing of the results.

Replicability requires replicable behavior of all building blocks of the experiment, for which the setup of particular tests is recommended. Commonly, three categories of tests are considered:
Unit tests, examining a small section of the source code; integration tests, checking a major component of the source code; and system tests, assessing the whole project \cite[Chapter~3]{swebok}.
Tests usually involve a comparison of the computed to analytical results, statistically significant sampling or the conformance to an accepted benchmark problem.

\subsection{Reproducibility Requirement: Extensive Documentation}\label{sec:reprod-req}
To enable the reproducibility of a CBEx, a sufficiently detailed description of the algorithms, implementation, test setup, and parameters needs to be provided.
Here, sufficiency is achieved if the documentation contains all information needed to setup and to run the experiment by a different researcher in a comparable environment.

However, to reproduce a CBEx in a different environment, a documentation of the utilized hardware and software is also needed.
An essential part of this environment documentation is the listing of other software packages required to perform the CBEx. 
Documenting these dependencies includes all software, which is not available in a commonly assumed environment with employed variant and version and allows to set up the same or at least similar software stack.

Depending on the programming language in which the considered CBEx is encoded, different types of dependencies arise.
A compiled language requires a compiler and linked libraries to generate an executable file embodying the program computing the results.
The variant of the compiler and its version as well as the variants of (statically and dynamically linked) libraries with their versions make up the associated dependencies.
Furthermore, a build system, which organizes the compilation and linking may be used and constitute a dependency.
An interpreted language requires an interpreter, which parses and executes the source during its runtime.
In this case, typical dependencies are the variant of the interpreter in a specific version as well as depending toolboxes with versions.

\subsection{Reproducibility Recommendation: Availability}
The \textbf{availability} of the source code archive is highly recommended for reproducibility because of two main reasons.
First, the code itself may serve as documentation of the experiment.
Second, the code may be used to realize the actual reproduction.

Therefore, the availability of the source code archive from a stable location is vitally important.
A location can be considered stable if its main purpose is storing data. 
This does not imply lasting availability, hence a second backup location is commendable.

The classic method of providing source code access is the bundling with the publication by including the source code archive as supplemental material.
This affiliates the code with the publication and is conveniently obtainable together with the publication itself.
Yet, a supplemental material section may not be available for all journals or may only accept certain file types (with a maximum file size).

Recently, software depots for scientific source code have been established.
For example, \textbf{RunMyCode}\footnote{\url{runmycode.org}} or \textbf{ResearchCompendia}\footnote{\url{researchcompendia.org}} are services storing source code archives and associating these to publications. 

Alternatively, the source code archive can be published separately through platforms such as \textbf{Zenodo}\footnote{\url{zenodo.org}} or \textbf{Figshare}\footnote{\url{figshare.com}}.
An advantage of this method is the assignment of a digital object identifier (DOI) for such a software publication, which can then be stated in the \hyperref[sc:cas]{Code Availability Section} of the associated publication.

As for the dependencies, reproducibility is not inhibited by closed-source software. However, a statement on the applicability of an open-source variant, if available, of those dependencies is suggested. In any case, those parts of the experiments that are not part of the source code, need to be documented as described in \href{sec:reprod-req}{Section~\ref*{sec:reprod-req}}.

\subsection{Reusability Requirement: Accessibility}\label{sc:reusereq}
A CBEx is reusable if it is accessible in a related or even different context.
\textbf{Accessibility} encompasses all means to (partially) apply the functionality of the original to another CBEx. 
The availability of source code fulfills the accessibility for reusability,
but also access to a compiled executable and library or a remote service is sufficient to comply.

\subsection{Reusability Recommendation: Modularity, Software Management \& Licensing}\label{sc:reuse2}
To be able to adapt a CBEx to differing environments and settings, the CBEx itself has to allow some parametrization to enable a certain configurability. 
Furthermore, \textbf{modularity}, the separation of experiment and method, enables the utilization of the method in other experiments or conducting the experiment with alternative methods.
A more fine-grained modularization can allow, in addition, the exchange of components from the method or experiment such as numerical solvers or service libraries.
Modularity necessitates a definition of interfaces which determine the communication between the interchangeable components.
The documentation of such an interface is essential for it to fulfill its purpose and involves e.g.\ a description of protocols, variables, types and function signatures with their arguments and return values.

Source code usually undergoes some evolution over time during which errors are fixed, and new features are introduced. 
Hence, \textbf{software management} methods, such as version control, are recommended for the organization of this development process.

A reusable software project is recommended to obey some versioning procedure.
A version scheme allows a unique identification of different chronological stages of the project.
Usually, such a version consists of at least two numbers delimited by a dot, describing the major and minor iteration of changes.
More fine-grained versioning can be applied with further numbers.
A release of a new version can be fixed by assigning a DOI.

To record the evolution of the source code a version control system, such as \texttt{git}, \texttt{mercurial} or \texttt{bazaar}, is an important tool.
A version control system tracks changes for each controlled file and allows a well-defined collaborative work on the source files.
The set of all files under version control makes a repository, a set of changes to a single or multiple files constitute a revision of the repository, and a set of revisions defines a new version.
A history of the revisions can also augment the documentation of the CBEx if the changes are recorded with comprehensive descriptions.

A license assigned to the source code archive,
which governs the rights and duties associated with its use and reuse as well as indicating copyrights, is practically necessary for reusability.
If an open-source license is selected, certain characteristics should be considered:
The license should be approved by the Open-Source-Initiative\footnote{\url{opensource.org}} and the Free-Software-Foundation\footnote{\url{fsf.org}} as well as being compatible with the GNU-General-Public-License\footnote{\url{opensource.org/licenses/gpl-license}}.
Generally, a central requirement for scientific software should be an attribution clause requiring the future inclusion of the copyright information, which usually notes authors and contributors.
A \emph{non-permissive} license may inhibit the reusability of the software in non-open projects, cf. \cite{stodden09}.
To select a license, the service \textbf{Choose-A-License}\footnote{\url{choosealicense.com}} can be of help,
and for an explanation of the selected license, a service like \textbf{tl;dr Legal}\footnote{\url{tldrlegal.com}} provides short summaries of the license's legal implications.

\section{Basic Documentation}\label{sc:definfo}
In terms of research software, it is important that the accompanying documentation enables usage and reproducibility of results.
To this end, certain information on the tested hardware and software should be documented.
Following, a basic form of documentation is proposed, which includes the essential information to facilitate RRR.

A simple form of documentation is providing basic information in plain text files.
These should be sequential files containing only printable ASCII characters
\cite{iso646} and consequently using a US-ASCII file encoding.
If it is necessary to also use non-ASCII characters, a modern encoding with good cross-platform support, like UTF-8, should be used.  
Recently, these text files have been decorated with \textbf{commonmark}\footnote{\url{commonmark.org}} mark-down code\footnote{Usually indicated by the file extension \texttt{.md}.},
which rather improves readability then inhibiting it and are considered an unofficial standard due to the widespread use for example by \textbf{github}.
Since typically scientific publications are composed in the English language, so should be these text file.

Certain default filenames are established to indicate the file's contents, such as \texttt{README}, \texttt{LICENSE}, \texttt{AUTHORS} and \texttt{CHANGELOG}.
Additionally, further files of relevance to the academic environment have been suggested such as \texttt{CITATION} and \texttt{CODE}.
This work proposes two more files, namely \texttt{RUNME} and \texttt{DEPENDENCIES} to facilitate replicability.

\subsection{README}
The bare minimum of any code package, source code repository or source code archive should be a \texttt{README} file.
To uniquely identify this text file it should state the name of the associated software project along with its version and the release date.
Normally, also a brief description of the package functionality and its contents are expected.

Often, the \texttt{README} file also includes a manual for the compilation or installation of the project.
In the case that these procedures are more elaborate, a separate \texttt{INSTALL} file can be used and referenced inside the \texttt{README}.
The same holds for the authors and contributors to the project,
which can be listed in the \texttt{README} or in an additional \texttt{AUTHORS} file.
Relevant information for the \texttt{README} includes a project website, a (stable) download location, contact information and sample usage (for example referencing the \texttt{RUNME} file) of the associated software.
Furthermore, the license and the \texttt{LICENSE} file\footnote{The \texttt{LICENSE} file holds the full license text the copyright holders and the release year}, a record of the history of changes in the \texttt{CHANGELOG} file, a set of frequently asked questions in a \texttt{FAQ} file and a documentation can be referenced.

In the case that the replicability of an experiment is targeted, the specifically used software stack and hardware environment should be documented,
as well as all configurations, parameters and arguments defining the CBEx.
For reproducibility, related publications should be cited, and for reusability, links to technical documentation, e.g. interfaces, or a version control repository could be listed.
Generally, a \texttt{README} file can also act as a table of contents to the remaining files associated with the source code archive.

Preferably, the \texttt{README} presents the necessary information to start using the software in a quick and comprehensive way.
Therefore, the general recommendation is to make it as detailed as necessary while at the same time keeping it as brief as possible.
For in-depth discussions of the further details, a reference to the actual software documentation should be preferred.

\subsection{RUNME}
To facilitate replicability, an additional file called \texttt{RUNME} is proposed in this work,
and lists the steps required to replicate results.
This can be an executable script file, which upon execution automatically performs all steps necessary to replicate the results of an associated publication.
In case, multiple environments are supported, the respective environment can be highlighted by a file extension, for example \texttt{RUNME.linux} or \texttt{RUNME.win}.
Alternatively, the \texttt{RUNME} file can describe these stages in pseudo-code or, in general, not machine readable language.

\subsection{CITATION}
The concept of a \texttt{CITATION} file has first been used by the \textsc{R-project} \cite{rproject} and has also been adapted by \textsc{GNU Octave} \cite{octave}.
This file contains information on how to cite the associated software project in other works. 
Besides a sample citation, a suggested \texttt{BibTeX} code is often provided in this file.

\subsection{DEPENDENCIES}
Modern software stacks encompass multiple layers of intermediary software on which a project may depend upon.
To be able to build and use a provided source code package such dependencies must be locally available.
For projects with few dependencies, it is sufficient to list those in the \texttt{README} file,
yet for projects with many dependencies it is suggested to include a \texttt{DEPENDENCIES} file that lists these necessary (third-party) software components including the required version.
Dependencies encompass, but are not limited to: runtime environments, libraries, toolboxes, source code archives or executable files.

\subsection{CODE}
The purpose of the \texttt{CODE} file is the listing of key meta-data on the associated software project.
Initially, the idea of bundling code meta-data was proposed in \cite{smith14} and formalized in \cite{katz15}.
The main intended purpose of this proposal was the assignment of transitive credit in software stacks utilized for scientific work.  
In publications, about a software project this meta-data also helps as a unique identification,
as for example in the \textit{SoftwareX} journal\footnote{\url{www.journals.elsevier.com/softwarex}}.
Another important reason for code meta-data is the classification and organization of scientific software, which facilitates reproducibility and reusability.
This information could and should also be enclosed in the \texttt{README} file,
yet the focused \texttt{CODE} file is machine-readable and allows automatically generated directories. 

Various file formats to encode this meta-data are surmisable.
Among others there are:
\texttt{ini} (Initialization File),
\texttt{xml} (Extensible Markup Language),
\texttt{yaml} (YAML Ain't Markup Language) and
\texttt{json} (Javascript Object Notation), which is suggested in \cite{smith14,katz15}.
Basic requirements for such a file are a plain text encoding and a human readable formatting.
Additionally, a simple syntax\footnote{This is understood as a small set of rules.} as well as the availability of parsing facilities should be considered.
Due to its renownedness and easy readability for human and machine, the authors suggest to use the \texttt{ini} file format, 
as the more elaborate grammars \texttt{xml}, \texttt{yaml} and \texttt{json} require sophisticated parsers.

There is no standard defining the \texttt{ini} format, yet its widespread use establishes a quasi-standard:
Each line in an \texttt{ini} file holds a single key-value pair, which is delimited by a colon.
The other formats also provide hierarchies for its components, which allow nesting of fields,
for example grouping an author's properties under a common author key,
but these hierarchies introduce an impediment for the automatic parsing of contents.
To resolve the former example of multiple authors, in the case of the \texttt{ini} file a comma separated list can be used as the value.

Due to the wide range of possible meta-data across the sciences utilizing software,
no one-size-fits-all list of keywords is given, but a list of suggestions which applies to most research software projects.

\begin{itemize}

 \item \textbf{name} The primary identifier of the software project.

 \item \textbf{shortname} An alias or the name of the main executable.

 \item \textbf{version} A unique state of the project, usually symbolized by numbers separated by decimal points indicating the major and minor revisions. 

 \item \textbf{release-date} The date this version has been released written in the ISO-8601 international format: \texttt{YYYY-MM-DD} \cite{iso8601}.

 \item \textbf{doi} A digital object identifier fixing a software release at a stable location.

 \item \textbf{authors} The list of authors.

 \item \textbf{orcids} The list of ORCID\footnote{\url{orcid.org}} identifiers corresponding to the list of authors.

 \item \textbf{topic} A basic categorization\footnote{For example category classifications such as MSC ( \url{msc2010.org} ), ACM (~\url{www.acm.org/about/class} ) or PACS ( \url{www.aip.org/publishing/pacs} ) may be used.} of the project. 

 \item \textbf{type} The type of software, for example a program, library or toolbox.

 \item \textbf{license} The license under which the software is released.

 \item \textbf{license-type} Distinguishes between open and propriety licenses.

 \item \textbf{repository} The link to project's source code repository.

 \item \textbf{repository-type} The type of version control software of this repository.

 \item \textbf{languages} This field is supposed to contain a comma separated list of utilized programming languages in the software project.
                          For larger projects a naming of the major languages will be sufficient.
                          Since programming languages evolve over time, a version or standard of the employed language or dialect should also be provided.

 \item \textbf{dependencies} A list of software required to use the project, such as libraries, toolboxes and runtimes.

 \item \textbf{systems} A list of compatible operating systems or computational environments.

 \item \textbf{website} If the CBEx is part of an enclosing research software project and has a website, the URL (Uniform Resource Locator) can be provided in this field to guide users to the available resources.

 \item \textbf{keywords} A list of descriptive terms.

\end{itemize}

~

An example of such a code meta data \texttt{ini}-file, from \texttt{emgr} - the empirical gramian framework \cite{emgr}, is shown in \href{fig:codemeta}{Figure~\ref*{fig:codemeta}}.

\begin{figure}[h!]
\begin{framed}
\texttt{name: Empirical Gramian Framework\\
shortname: emgr \\
version: 3.9 \\
release-date: 2016-02-25 \\
doi: 10.5281/zenodo.46523 \\
authors: Christian Himpe \\
orcids: 0000-0003-2194-6754 \\
topic: Model Reduction \\
type: Toolbox \\
license: 2-Clause BSD \\
license-type: Open \\
repository: github.com/gramian/emgr \\
repository-type: git \\
languages: Matlab \\
dependencies: GNU Octave >= 3.8, MATLAB >= 2011b \\
systems: Linux, Windows \\
website: gramian.de \\
keywords: empirical gramians, cross gramian, combined reduction
 }
\end{framed}
 \caption{Sample \texttt{CODE} \texttt{ini}-file for the empirical gramian framework.}
 \label{fig:codemeta}
\end{figure}

\subsection{Source Code File Headers}
Apart from the text files enclosed with the project, every source code file should state in its first lines, the so-called header:
\begin{enumerate}
 \item the associated project,
 \item the authors and contributors,
 \item and the purpose of the file.
\end{enumerate}
This establishes the affiliation of this source file to the project.
The header can optionally also include license and version information.
Additionally, this file header can hold citations to works used to compose the following source code or keywords categorizing the contents.

%
%
%

\section{A Practical Example}\label{sc:min_max_example}

\begin{figure}[htp]
  \centering
   \includegraphics{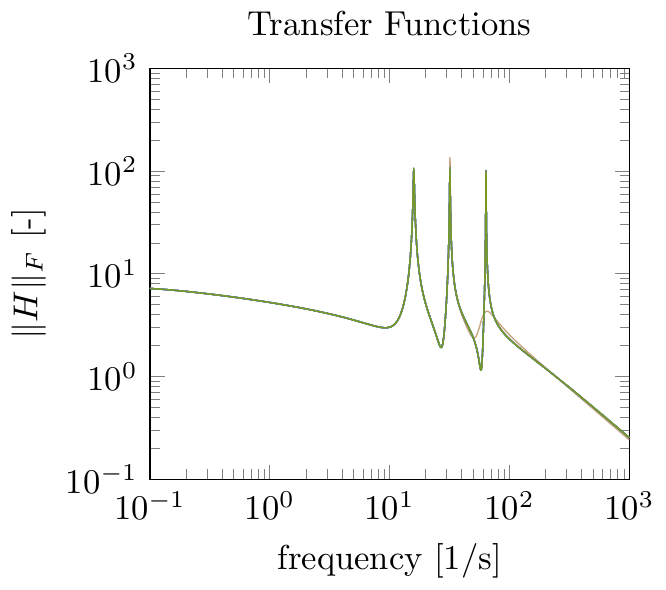}
   \includegraphics{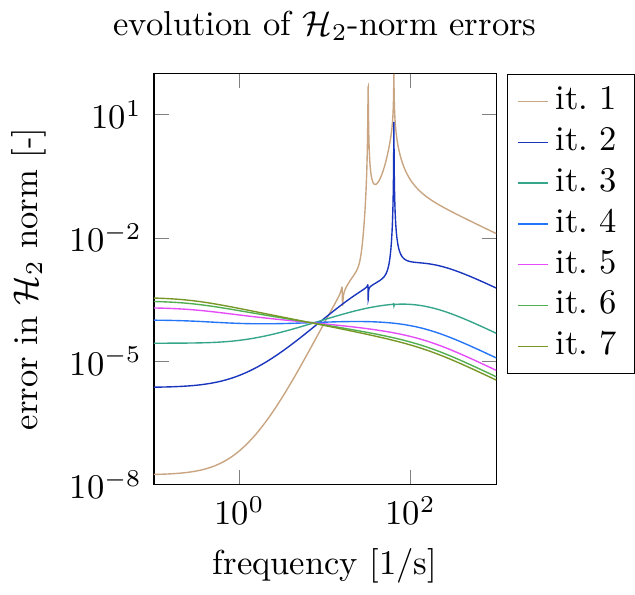}\\
   \includegraphics{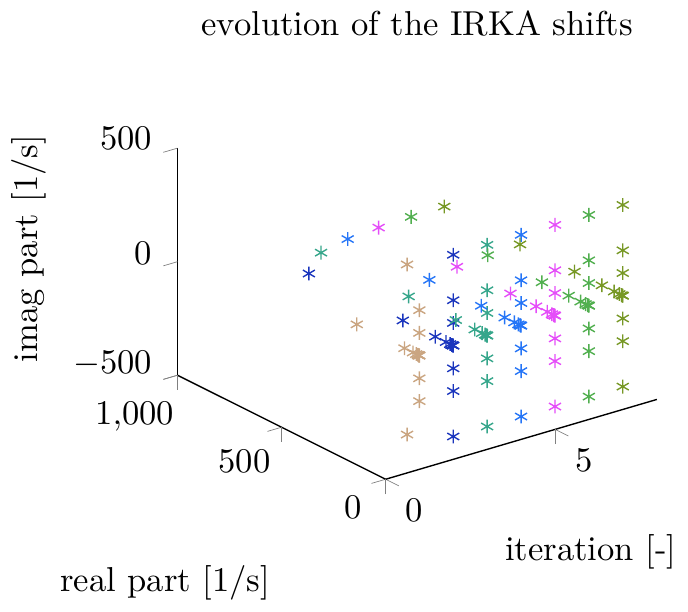}
   \includegraphics{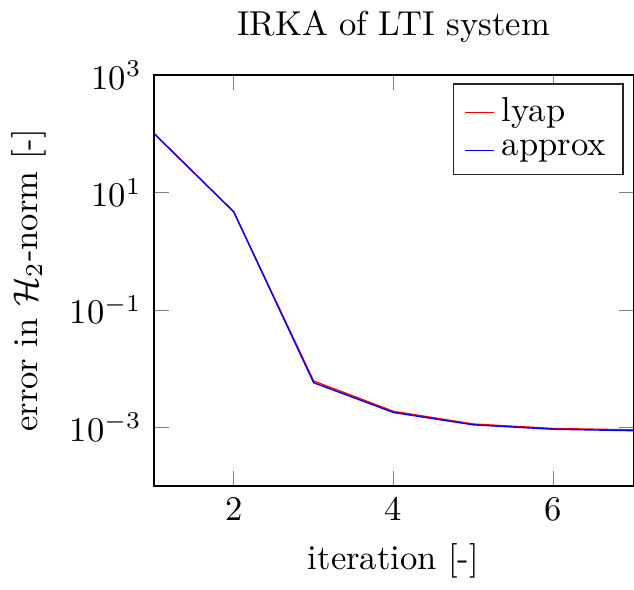}
  \caption{Example IRKA results for the FOM model by Penzl and reduced order
    10.} 
  \label{fig:IRKA}
\end{figure}

In this section, we discuss a very rudimentary and simple implementation of the iteratively corrected rational Krylov algorithm for $\mathcal{H}_2$ model reduction proposed by Gugercin, Antoulas and Beattie~\cite{GugercinAntoulasBeattie08}.
The implementation of the algorithm was made as an exercise in a lecture about model reduction.
The common denominator of the authors is the fact that their research is within the area of model order reduction. 
But, their backgrounds, scientific computing, mathematics, control or engineering, is different.
Nevertheless, in our opinion, the sharing of code, good documentation, and modular programs which can be reused is essential for the further success of model order reduction. 
The intention of the best practice example is exemplary to show the files and rules for good CBEx’s. The example serves as a template for other research.
During implementation, we particularly paid attention to follow the guidelines given in this work. In a first step, the IRKA algorithm \cite{GugercinAntoulasBeattie08} is chosen because the algorithm is widely used, heavily cited algorithm but also has a well-documented examples section where the numerical experiments used to verify the behavior of the algorithm are described including the model.
Also, the outcome of the algorithm is for many examples deterministic therefore replicability of the results of \cite{GugercinAntoulasBeattie08} is achieved. The minimum requirement for replicability is the basic documentation, which documents the RUNME.m file and every single function. Two example files are given. In the first example, RUNME.m, the IRKA algorithm automatically produces the figures shown in \href{fig:IRKA}{Figure~\ref*{fig:IRKA}}. The second example file, EXAMPLES.m, can be used to test the algorithm with different test examples and is used to test the algorithm on various system architectures with different programs and different program version. Documentation in the header, which architectures and programs work with the algorithm and the test examples, is recommended. Furthermore, standardized benchmark examples, e.g.\ from the Oberwolfach Benchmark Collection\footnote{\url{portal.uni-freiburg.de/imteksimulation/downloads/benchmark}} are used to allow reproducibility of the results for other users. 
Finally, to demonstrate the advantages of reusability part of the implementation is based on the work of Panzer~\cite{Panzer14}. Since the source code of Panzer~\cite{Panzer14} is published under an open-source license, a reuse of his work of is possible. We can modify and use the code for our own purpose. Consequently, for a further reuse of the source code, this implementation is also published under a public license. The code was made public via a GitLab archive\footnote{\url{gitlab.mpi-magdeburg.mpg.de/saak/best\_practice\_IRKA.git}} and uniquely identified and archived via a Zenodo entry with a valid DOI~\cite{FehrSaak16}, the availability of the source code is depicted in our Code Availability section below. Nevertheless to show the possibility to combine open source code with closed source code the function \textit{calculateFrequencyResponse.p} is given in a p-coded version, which is obfuscated to protect intellectual properties.

The results shown in \href{fig:IRKA}{Figure~\ref*{fig:IRKA}} use Penzl's FOM
benchmark example (see e.g., \cite[Section C.3.1]{Pen00a}) and apply our
implementation of the method from \cite{GugercinAntoulasBeattie08}. In the
reported test the initial shift parameters and the reduced order have been
chosen such that the progress of the IRKA iteration becomes nicely
visible. Larger reduced orders would allow for smaller error norms, while more clever choices of the initial shift could lead to less overall iterations. Both are however beyond the scope of this presentation.
\section{Closing Remarks}
In this contribution the notions of replicability, reproducibility and reusability are discussed and classified by requirements and recommendations.
The issue of code availability and the implied reflection on the artifacts of associated CBEx is exemplified,
and simple formats of documentation and meta-data provisioning are described.

The proposed best practices in this work improve scientific validity of CBEx, but also aim to spark a discussion on RRR in this context.
And by no means are the suggested techniques to be understood as a strict rulebook with everlasting validity.
The authors emphasize that the proposed practices, which are based on practical experience and standards as well as on general considerations of abstract concepts, are subject to change over time.
Nonetheless, the herein demonstrated strategies do enhance replicability, reproducibility \& reusability and thus, also in the absence of other general solutions or approaches, merit their consideration for scientific CBEx in general and numerical CBEx in particular.

\section*{Code Availability} 
~\\
\begin{minipage}{\linewidth} 
	\begin{framed}
		The source code of the implementations used to compute the presented results can be obtained from: \\
		\begin{center}
			\href{http://dx.doi.org/10.5281/zenodo.55297}{doi:10.5281/zenodo.55297} and is authored by: J\"org Fehr and Jens Saak
		\end{center}
		~\\ Please contact J\"org Fehr and Jens Saak for licensing information
	\end{framed}
\end{minipage}

\section*{Acknowledgements}
This work was supported by the Deutsche Forschungsgemeinschaft, DFG EXC 1003 Cells in Motion -- Cluster of Excellence, M\"unster, the Center for Developing Mathematics in Interaction, DEMAIN, M\"unster, Germany,
and the Deutsche Forschungsgemeinschaft, DFG EXC 310/1 Simulation Technology at the University of Stuttgart.

\section*{Conflict of Interest}
All authors declare no conflicts of interest in this paper.

\bibliographystyle{plain}
\bibliography{references}
\end{document}